\documentstyle[preprint,eqsecnum,aps]{revtex}
\begin{document}
\draft
\def \beq{\begin{equation}}
\def \eeq{\end{equation}}
\def \beqarr{\begin{eqnarray}}
\def \eeqarr{\end{eqnarray}}

\title{A Field Theory for the Read Operator}
\author{R. Rajaraman\cite{byline1}}

\address{School of Physical Sciences \\
Jawaharlal Nehru University\\ New Delhi 110067, \ INDIA\\ }
\author{S. L. Sondhi\cite{byline}}
\address{Department of Physics\\
University of Illinois at Urbana-Champaign\\ Urbana, IL
61801-3080, U.S.A\\ }
\date{November 19, 1995}
\maketitle
\begin{abstract}
We introduce a new field theory for studying quantum Hall systems.
The quantum field is a modified version of the bosonic operator
introduced by Read. In contrast to Read's original work we do {\em not}
work in the lowest Landau level alone, and this leads to a much simpler
formalism. We identify an appropriate canonical conjugate field, and
write a Hamiltonian that governs the exact dynamics of our bosonic field
operators.  We describe a Lagrangian formalism, derive the equations
of motion for the fields and present a family of mean-field solutions.
Finally, we show that these mean field solutions are precisely the Laughlin
states. We do not, in this work, address the treatment of fluctuations.

\end{abstract}
\section{Introduction}

Much of our current understanding of the novel behavior in the
quantum Hall (QH) regime relies upon microscopic, first quantized,
wavefunctions for ideal QH states. The nature of the ordering in
these states was first elucidated by Girvin and MacDonald \cite{GM}
who showed
that the Laughlin wavefunctions exhibit a non-trivial form of
off-diagonal long range order. More precisely, they showed that
there exists a composite operator of the fundamental Fermi
fields that obeys Bose statistics and whose off-diagonal density matrix
is algebraically long ranged in the Laughlin states \cite{fn5}.
This was an extremely
important observation for it opened up the possibility of a Landau-Ginzburg
description of the QHE in terms of an order parameter field, thereby
bringing it within reach of more systematic computations. Subsequently
Zhang, Hansson and Kivelson (ZHK) \cite{ZHK} and Read \cite{Read}
proceeded to construct such Landau-Ginzburg theories.

The approach taken by ZHK builds directly on the observation of GM and
consists of reformulating the dynamics entirely in terms of the bosonic
operator. An exact transformation allowed them to rewrite the partition
function of two-dimensional electrons placed in an external magnetic field
in terms of a set of bosons that continue to see the external magnetic field
but also interact among themselves through a statistical or Chern-Simons
gauge field that exactly reproduces the Fermi statistics of the original
problem. In the new variables a novel mean-field approximation becomes
possible, in which the statistical gauge field cancels part of the external
magnetic field. In fact at the ``magic'' fractions in the QHE the cancellation
is exact and the bosons are able to condense. So in the bosonic formulation,
QH states emerge as condensates of certain composite bosons.

While this mean-field picture is ``morally'' correct, it isn't quite right
in its details. Recall that we mentioned that GM found only algebraic
order for the bosons; that is inconsistent with the true long range order
present in the ZHK mean-field state. Fixing this requires that we include
corrections from gaussian fluctuations (RPA) \cite{zhang}. The problem
here is that the ZHK field theory takes no advantage of the Landau level
structure that the magnetic field induces; indeed, the mean field wavefunction
has extensive content in high Landau levels. Consequently, the Landau level
structure has to emerge {\em dynamically} and starts to do so only
at the RPA level where it can be shown that the asymptotic form of the
wavefunction is now of the Laughlin form \cite{zhang,lopez}.

In contrast, Read was concerned with obtaining a true Landau-Ginzburg
description in the sense of writing down classical equations of motion
for an order parameter that really has an expectation value. To this
end he worked with a bosonic operator that takes advantage of the
Landau level structure and whose off-diagonal density matrix is truly
long ranged in the Laughlin states. Equivalently, the Laughlin states
are ideal condensates of the Read bosons. He then derived a rather complicated
effective action that described the dynamics of this operator for small
deviations from the Laughlin states that do not violate the constraint
of confining the electrons to the lowest Landau level.

While Read's approach builds in more of the physics of high magnetic
fields (i.e. the lowest Landau level constraint) it has proven to be
unwieldy for actual calculations. By contrast, the ZHK scheme both
in its original form and in a fermionic version \cite{lopez2} has
proven quite useful, e.g. in work on the phase diagram of QH systems
\cite{KLZ} and perhaps most notably in the work of Halperin, Lee
and Read himself \cite{HLR} on the $\nu=1/2$ problem. Nevertheless, it
suffers from various
problems stemming from the uncontrolled inclusion of higher Landau level
processes, which brings us to our objective in this paper.
(We should note that the Read operator does have the advantage, over
its GM/ZHK analog, of being easier to compute in numerical work.
This was exploited by Rezayi and Haldane \cite{RH1} in their
demonstration that the Read operator fails to condense when the
Laughlin states are destabilized by varying the inter-electron
interaction, and more recently by Sondhi and Gelfand \cite{SG} in
establishing that the condensation occurs everywhere in the QH phases
that derive from the Laughlin states in the presence of disorder.)

In this work we construct an exact quantum field theory for a slightly modified
version of the Read operator  in the full Hilbert
space of the system. It combines the virtues of the ZHK and Read formulations
in that, 1) the equations of motion
capture the full dynamics and are not restricted to small fluctuations
about the Laughlin states, 2) they are far simpler than if the lowest Landau
level constraint were imposed explicitly, and 3) yet the mean field solutions
are still precisely the Laughlin states.
 Our hope is that a calculation at the RPA level in
this formulation will yield useful information on mode spectra and corrections
to the Laughlin wavefunctions beyond what can be gleaned from a comparable
calculation in the ZHK approach. However, we warn the reader that we
have not yet solved the problem of actually systematizing such calculations
and detail the relevant difficulties in Section \ref{summary}.

In the remainder of the paper we describe the operator algebra for
the bosonic field theory, the derivation of the Hamiltonian, a Lagrangian
formulation and its mean-field solution and the equivalence of the
mean-field solution to the Laughlin wavefunctions. We close with a
summary.

\section{The Bosonic Formulation}

We consider a set of two-dimensional electrons of mass $\mu$ and charge $+e$,
placed in a uniform transverse magnetic field of strength $B$.
We take the electrons to be spinless for simplicity and leave
unspecified their interaction potential $V$ as well as the
scalar potential $A_0$ that represents any uniform and/or random
electric fields in the problem. The field theoretic (second quantized)
Hamiltonian that describes our system is,
\beq
H  =  \int d^{2}x\, \bigg[\Psi^{\dagger}({\vec x})\,
\bigg({-\hbar ^{2} \over 2\mu} {\vec D}^{2}   + eA_0 \bigg)\,
 \Psi ({\vec x})  \ \bigg]    +
  {1 \over 2} \int \int d^2 x\, d^2 x'  \,
\delta\rho({\vec x})\,  V(\vec {x}-\vec {x}') \, \delta \rho
({\vec x}') \ . \label{Ham}
\eeq
Here $\Psi({\vec x})$ is the full electron field operator and not merely
its lowest Landau level piece, $ {\vec D} \equiv {\vec \nabla}  -  i
{e \over {\hbar c}} {\vec A}$ is the covariant derivative inclusive of the
vector potential of the uniform field ($B = \nabla \times {\vec A}({\vec x}$))
and $\rho ({\vec x}) \ \equiv  \ \Psi^{\dagger}({\vec x}) \Psi({\vec x})$
is the density operator whose deviation from its mean value $\bar \rho$ is
$ \delta \rho ({\vec x})$.
In addition to specifying $H$ we need to require that $\Psi$
is a Fermi field, i.e. it obeys the equal-time anticommutation relations,
\beqarr
\{ \Psi({\vec x}) , \Psi({\vec x}') \}  &=& \{ \Psi^{\dagger} ({\vec x}) ,
 \Psi^{\dagger}({\vec x}') \}  =  0 \nonumber \\
\{\Psi({\vec x}),\Psi^{\dagger}({\vec x}') \} &=&
\delta^{(2)} (\vec {x}-\vec{x}') \ .
\eeqarr
In  order to recast this as a bosonic problem we need a set of bosonic
operators which we construct in the next section.

\subsection{Bosonic Operators}

Consider the pair of operators $\Phi({\vec x})$ and $\Pi({\vec x})$,
defined by
\beqarr
\Phi({\vec x}) &\equiv&   e^{ -J({\vec x})} \Psi({\vec x})
 \nonumber \\
\Pi({\vec x}) &\equiv& \Psi^{\dagger}({\vec x}) e^{ J({\vec x})},
\label{fipi}
\eeqarr
where,
\beq J({\vec x}) \equiv  m\int d^{2}x'\,
[\rho ({\vec x}')  \log(z-z')]
   -  {\mid z \mid ^2 \over 4l^2}  , \label{defJ}
\eeq
where m is an odd integer,  $z \equiv x_1+ix_2$  is the complex
coordinate on the plane and $l  =  \sqrt{{\hbar c \over eB} }$ is the
magnetic length. Evidently, $\Phi$ and $\Pi$ are not hermitian conjugates
as $J$ has both hermitian and anti-hermitian pieces; in fact,
\beq
\Pi({\vec x}) = \Phi^\dagger({\vec x})
e^{ J({\vec x}) + J^\dagger({\vec x})} \ .
\label{piphirelation}
\eeq
Nevertheless, as we now show, they are canonically conjugate Bose
fields.

To this end note that the only operator appearing in J({$\vec x$}) is
the electron density
$\rho ({\vec x}) = \Psi^{\dagger}({\vec x}) \Psi({\vec x})$, which obeys
the commutation relation,
\beq
\big[ \rho({\vec x}), \Psi({\vec x}')\big]  =  -\Psi({\vec x})\,
\delta^2 ({\vec x}-{\vec x}') \ .
\eeq
Using this one can obtain the following identities:
\beqarr
e^{-J({\vec x})}\, \Psi({\vec x}') &=& (z-z')^m \,  \Psi({\vec x}')\,
e^{-J({\vec x})} \nonumber  \\
\Psi^\dagger({\vec x}')\, e^{-J({\vec x})} &=& (z-z')^m \, e^{-J({\vec x})}
\, \Psi^\dagger({\vec x}')
\ . \label{ident}
\eeqarr
It is then straightforward to verify using (\ref{ident}) that
\beq
[\Phi({\vec x}), \Phi({\vec x}')]  =  [\Pi({\vec x}) ,
\Pi({\vec x}')]  =  0
\label{bcomm1}
\eeq
while
\beq
[\Phi({\vec x}) ,\Pi({\vec x}')]  =  \delta^2 ({\vec x}-{\vec x}') \ .
\label{bcomm2}
\eeq
Thus, despite the presence of non-unitary factors in their definition in
Eq. (\ref{fipi}),
the fields $\Phi$ and $\Pi$ form a pair of mutually canonical Bose fields.
However, in contrast to standard charged scalar field theories, here $\Pi$
is not equal to $\Phi^{\dagger}$, instead they obey the more complicated
relation in Eq. (\ref{piphirelation}). This fact, a consequence of the
non-unitary transformation in (\ref{fipi}), has to be borne in mind in doing
manipulations with our theory.

Nevertheless, notice   that the fermion density $\rho$, when
written in terms of $\Pi$ and $\Phi$ still has the standard bosonic form
\beq
\rho({\vec x})  =  \Psi^{\dagger}({\vec x}) \, \Psi({\vec x})  =
\Pi({\vec x}) \Phi({\vec x}) \ .  \label{rhobos}
\eeq
Thus, if $N \equiv \int d^2x\, \rho$ is the number operator, then
\beq
\big[ N ,  \Pi({\vec x}) \big]  =  \Pi({\vec x}), \label{Npi}
\eeq
i.e. the operator $\Pi({\vec x})$ creats one extra composite boson,
and the number
of composite bosons is the same as the number of the original fermions.

Readers familiar with the work of ZHK will recognize that in the
corresponding construction of a Bose field in their approach the
operator $J$ was chosen to contain only the phase of $(z-z')$, i.e.
$ {\rm Im} \log (z-z')$ with the consequence that their $\Pi = \Phi^\dagger$.
In Read's work the creation operator for the bosons differs from our
$\Pi$ only in that he did not include the gaussian factor
$e^{-{\mid z \mid ^2 \over 4l^2}}$ in its definition; hence our bosons
are essentially the same. The difference between Read's work and ours
arises in that we also construct the canonical conjugate of $\Pi$ and
that, as we show next, allows us to write down a canonical quantum field
theory of bosons by changing variables from $\Psi$ and $\Psi^{\dagger}$ to
$\Phi$ and $\Pi$.

\subsection{The Hamiltonian}

Consider the action of the covariant derivative on the electron field.
We have,
\beqarr  {\vec D} \Psi(x) \ &=& \ {\vec D} \big( e^{J({\vec x})} \Phi
 ({\vec x}) \big) \nonumber \\
 \ &=& \ \big({\vec \nabla} \ - \ i {e \over {\hbar c}} {\vec A}({\vec x})\big)
  \  \big( e^{J({\vec x})} \Phi ({\vec x}) \big) \nonumber \\
  &=& \ e^{J({\vec x})} \ \ \big( {\vec \nabla} \ - \ i {e \over {\hbar c}}
   {\vec A}({\vec x}) \ + \ {\vec \nabla} J({\vec x}) \big )
    \ \Phi({\vec x}) \nonumber \\
 \   &=&  \ e^{J({\vec x})} \big ( {\vec D} -{ie \over {\hbar c}}
{\vec v}({\vec x})   \big) \  \Phi({\vec x}) \eeqarr
where,
\beq
{\vec v}({\vec x}) \equiv  i{\hbar c \over e}
{\vec \nabla}J({\vec x})  \ .
\label{defa}
\eeq
Hence,
\beq
D^2 \Psi  =  e^J \, \big( {\vec D}  -  i{e \over \hbar c} {\vec v}
\big)^2  \Phi
\eeq
Inserting this into the starting Hamiltonian (\ref{Ham} ), and using
Eqs. (\ref{fipi}) and (\ref{rhobos}) we get,
\beqarr
H  &=& \int  d^2x  \, \bigg[\Pi ({\vec x}) \,
\bigg( {-\hbar ^{2} \over 2\mu}   \big({\vec \nabla}  -
 {ie \over \hbar c} \ (\vec {A} + \vec {v})
\big)^{2}  +  eA_0 \bigg) \Phi ({\vec x}) \bigg] \ \nonumber \\
& +&  {1 \over 2}  \int  \int \, d^2x \, d^2 x'
\,\delta\rho({\vec x})\, \ V({\vec x}-{\vec x}') \, \delta\rho ({\vec x}')
\label{HB}
\eeqarr
This Hamiltonian, the auxilliary definitions (\ref{rhobos}) and
(\ref{defa}) and the commutators (\ref{bcomm1}, \ref{bcomm2}) together define
a purely
bosonic problem that is fully equivalent to our original fermion
problem \cite{fn1}.

The vector field ${\vec v}$ appearing in \ref{HB} above is constrained
in terms of the density by Eq. (\ref{defa}),
where J({$\vec x$}) is defined in (\ref{defJ}). Since this J({$\vec x$})
involves
more than just the phase of $(z-z')$, this field ${\vec v}$ is not
the familiar
statistical Chern-Simon gauge field ${\vec a}_{cs}$ used, for instance,
in \cite{ZHK}. Because J({$\vec x$}) has real parts, $ {\vec v }$
is a complex vector field. However, we will see now that
${\vec v}$ is  simply related to $ {\vec a_{cs}}$.

In our notation the Chern-Simons field is defined as
\beq
{\vec a}_{cs}({\vec x})   =  {-m\hbar c \over e} {\vec \nabla}_x
\int \, d^2 x' \, \rho({\vec x}') \, {\rm Im}  \log (z-z')   \ ,
\eeq
or equivalently
\beq
b \equiv  \nabla \times  \ {\vec a}_{cs} \ = \ - \ m \phi_0 \rho
\label{defacs}
\eeq
where $\phi_0 \equiv {hc \over e }$ is the flux quantum.
Now, the function log z obeys the Cauchy-Riemann conditions away from
$ z \ = \ 0$, which can be written as
\beq
{\vec \nabla} ( {\rm Re} \log z)  =  {\vec \nabla} ( {\rm Im} \log z)
\times {\hat k}
\eeq
where  ${\hat k}$ is a unit vector perpendicular to the plane.
Using this we get,
\beqarr
{\vec v}({\vec x})  &=&  {i \hbar c \over e} {\vec \nabla}
 J({\vec x}) \nonumber \\
&=&  {i \hbar c \over e} \  {\vec \nabla}_x
m \int d^2 x' \, \big[\rho({\vec x}') ({\rm Re} \log (z-z')  +  i\,
{\rm Im} \log (z-z'))
\big] \ - \ {\mid z \mid ^2 \over {4 l^2}} \nonumber \\
 &=& {\vec a}_{cs}({\vec x})  +  i\,{\hat k} \times {\vec
a}_{cs}({\vec x})  -  {i \hbar c \over e} {{\vec x} \over 2l^2} \ .
\label{full a}
\eeqarr

Note that the last term in the above equation is just a
c-number term involving the coordinate vector $\vec x$. The
density dependent operator part of $\vec v$ is present {\em entirely}
through ${\vec a}_{cs}$.

\section{The Lagrangian and its Mean Field Solution}

In constructing a Lagrangian formulation it is very useful to implement
the constraint (\ref {defacs}), relating
${\vec a}_{cs}$ to the density $\rho$, by the usual device of introducing
a Lagrange multiplier field and recognizing the resulting gauge field
action as being the restriction of the Chern-Simons term to transverse
vector fields, ${\vec \nabla} \cdot {\vec a}_{cs} = 0$ \cite{ZHK}.
Thus the  Hamiltonian (\ref{HB}) along with the constraint (\ref{defacs})
will emerge from the following Lagrangian density:
\beq \ {\cal L} \ = \ \Pi \ (i\hbar \partial_t \ - \ e a_0  )
 \Phi \ - \
{e \over 2m \phi_0} \epsilon^{\mu \nu \sigma} a_{\mu} \partial_{\nu}
a_{\sigma} \  \ - \ \ H \label{lag} \eeq
where H is the bosonized Hamiltonian
in (\ref{HB}), and $a^{\mu} \ (\mu =0,1,2)$ is the 3-vector \ $(a_0,
{\vec a})$ and we have dropped the subscript on the Chern-Simons gauge
field.

There is, however, a subtlety in this procedure which does not arise
in ZHK's construction and has to do with the gauge invariance of the
resulting action and hence the freedom to pick gauges different from
transverse gauge. First, note that the action is manifestly invariant
with respect to gauge changes of the external field, i.e. the
transformations,
\beqarr
{\vec A} &\rightarrow & {\vec A} \ - \ {\hbar c \over e} {\vec \nabla} \Lambda
({\vec x},t) \nonumber \\
A_{0} &\rightarrow & A_{0} \ + \ {\hbar \over e} \partial_t \Lambda({\vec x},t)
\nonumber \\
\Phi \ &\rightarrow&  e^{- i\Lambda({\vec x},t)  } \ \Phi \nonumber \\
\Pi \ &\rightarrow &  e^{+ i\Lambda({\vec x},t)} \ \Pi
\ . \label{extgauge}
\eeqarr
However, the invariance with respect to gauge changes of the Chern-Simons
field is more restricted. Gauge transformations of the form (\ref{extgauge})
with $(A_0,{\vec A})$ replaced by $(a_0,{\vec a})$ leave the action invariant
only if $\Lambda({\vec x},t)$ is independent of ${\vec x}$, i.e. if they
do not involve the spatial gauge field at all. In addition, there is a class
of modified gauge transformations for the spatial components that do not
involve
the temporal component of the gauge field and have the following form.
Let $f(z) = u({\vec x}) + i w({\vec x})$
be an analytic function of $z$. Then the action is invariant under,
\beqarr
{\vec a} &\rightarrow & {\vec a} \ - \ {\hbar c \over e} {\vec \nabla} w
({\vec x}) \nonumber \\
\Phi \ &\rightarrow&  e^{- f(z)} \ \Phi \nonumber \\
\Pi \ &\rightarrow &  e^{ + f(z) } \ \Pi
\ .
\label{modgauge}
\eeqarr
(The variation of the gauge field implies that $J \rightarrow J +  f$ which
ensures that the constraint Eq. (\ref{piphirelation}) is preserved.)

As a consequence, the Chern-Simons field here is not really a gauge field
and by the Chern-Simons term we necessarily mean its restriction to
transverse gauge \cite{fn7}. This is not a practical issue in this
paper or in the computations we have in mind for they are typically
carried out in transverse gauge anyway. Nevertheless, the implications
of this feature of our theory, in particular the significance of the
modified gauge invariance (\ref{modgauge}) and its possible connection
to work on $W_\infty$ algebras \cite{winfty}, remain a subject for future work.

The field equations arising from the Lagrangian density (\ref{lag})
are a ``non-linear Schr\"odinger equation'' \cite{fn2},
\beqarr
\big(i\hbar \partial_t \ - \ e(a_0 \ + \ A_0)\big) \Phi({\vec x}) \ &=&
\ -{\hbar^2 \over 2\mu}
\bigg[{\vec \nabla} \ - \ {ie \over \hbar c} \bigg( {\vec A} \ + \ {\vec a}
 \ + \  i{\hat k} \times {\vec a} \ - \ {i\hbar c \over e} {{\vec x}
 \over 2l^2} \bigg) \bigg]^2 \Phi({\vec x}) \nonumber \\
& & + \left( \int \ d^2 x'  V({\vec x}-{\vec x}') \ \delta \rho ({\vec x}')
\right) \Phi({\vec x}) \ ,
\label{phieq}
\eeqarr
along with the modified Chern-Simons field-current identities,
\beq
\nabla \times {\vec a}  \  = \   -m \phi_{0} \Pi \ \Phi \eeq
\beq \hat{k} \times \left( - \partial_0 {\vec a} - {\vec \nabla} a_0 \right)
\ =  \ {m \phi_{0} \over c} \big( {\vec j}  \ - \ i{\hat k} \
\times \ {\vec j} \big ) \label{aeq} \eeq
  where,

 \beq {\vec j} \ = \  {\hbar \over 2 \mu i} \left[ \Pi\,({\vec {\cal D}}\,
  \Phi) - ({\vec {\cal D}}\,\Pi)\, \Phi \right] \eeq
 \beq {\vec {\cal D}} \ \equiv \  {\vec \nabla} \ - \ {ie \over \hbar c}
 \bigg( {\vec A} \ + \ {\vec a}
 \ + \  i{\hat k} \times {\vec a} \ - \ {i\hbar c \over e} {{\vec x}
 \over 2l^2} \bigg) \eeq
Although this current ${\vec j}$  does not look manifestly hermitian,
it is in fact just the usual hermitian electron current operator, as
can be verified by rewriting it in terms of the Fermi fields.

These equations have a simple mean field solution \cite{fn6} for the situation
where the external electric potential $A_0$ is absent and the uniform
magnetic field $B  = \nabla \times {\vec A}$ is chosen so that
the filling fraction is
\beq
\nu \ \equiv {\bar \rho \phi_0 \over B} \ = \ {1 \over m} \ ,
\eeq
where m is the odd integer in the fermion to boson transformation
function J defined in (\ref{defJ}). The solution describes a homogeneous
state and is given by the fields,
\beqarr
\Phi({\vec x}) &=& \Pi({\vec x}) \ = \ \sqrt{{\bar \rho}} \nonumber \\
{\vec a}_{\bar \rho}({\vec x}) &=&
{m \bar\rho \phi_0 \over 2} ( {\vec x} \times \hat{k})
\nonumber \\
a_0 &=& 0 \ .
\label{mfsoln}
\eeqarr
As the boson field has a uniform phase and non-vanishing amplitude
everywhere, this solution describes an ideal condensate of the composite
bosons \cite{fn3}.

In order to verify that this is indeed a solution, we begin by noting
that density $\rho = \Pi \Phi$ equals its mean value $\bar \rho$
everywhere. It follows that a constant $\Phi$ solves (\ref{phieq})
provided the gauge fields that enters the covariant derivatives vanish.
For the temporal gauge field this is trivially true. For the spatial
gauge field we note that,
\beqarr
{\vec a}_{\bar \rho}({\vec x})  &=&  -{1 \over 2}m \phi_0 \bar
\rho \, {\hat k}  \times  {\vec x} \nonumber \\
 &=&  -{1 \over 2} B  \, {\hat k} \times  {\vec x} \nonumber \\
 &=& -{\vec A},
\eeqarr
and hence the combination ${\vec a} + {\vec A}$ vanishes.
This condition for picking out uniform states, that the Chern-Simons field
at mean density cancels the external field ${\vec A}$, is already known
from \cite{ZHK}. But the statistical gauge field appearing in the
covariant derivative in
our Lagrangian and field equation (\ref{lag} and \ref{phieq}) is not just
${\vec A} \ + \ {\vec a}_{cs}$. It also contains imaginary pieces.
However,  we also have the additional the result that
\beqarr
{\hat k} \times \vec{a_{\bar \rho}}  &=&  { B \over 2}  {\vec x}
 \nonumber \\
&=&   {c \hbar \over e}  {{\vec x} \over 2l^2}
\eeqarr
This last equality tells us that the extra imaginary pieces of the
statistical gauge field, i.e. the third and fourth terms in Eq. (\ref{full a}),
also cancel one another.
Altogether, we have, for $\rho \ = \bar \rho$,
\beq
\vec A  + \vec {a_{\bar \rho}}  +i\, {\hat k}  \times
{\vec a_{\bar \rho}}
 - i\, {c \hbar \over e} \, {{\vec x} \over 2l^2}  =  0
\eeq
and hence the forms (\ref{mfsoln}) satisfy Eq. (\ref{phieq}).

It is also straightforward to verify that the forms (\ref{mfsoln}) solve the
field-current identity (3.5). Finally, readers concerned about
the consistency of the solutions for $\Phi$ and $\Pi$ should note
that for our solutions $J + J^\dagger =0$.

\section{The Mean Field Wavefunction}

We now show that the mean field solution of the last section is, in the
first quantized fermionic representation,
exactly the Laughlin state.
To see this note from Eq. (\ref{Npi}) that in our bosonized formulation,
an N-particle state is obtained by the action of N powers of $\Pi$ on the
vacuum. Hence the translationally invariant mean field state, where
all the bosons have condensed into the $k=0$ mode, has the (arbitrarily
normalized) form,
\beq
\vert  N \rangle_{MF}  =  {1 \over N!}\, \bigg[ \int d^2 x
\  \Pi({\vec x}) \bigg]^N \vert O \rangle
\eeq
where $ \vert O \rangle $ is the no particle (vacuum) state, and
$N,V \rightarrow \infty$ with ${N \over V} \ = \ \bar \rho$.
The first quantized {\em electron} wavefunction associated with this state is
\beqarr
\psi_{MF} \ ( {\vec y}_1,{\vec y}_2, \ ..... \ {\vec y}_N) \  \ &=& \
  \ \langle  O \vert \, \Psi({\vec y}_1) \ ....... \ \Psi({\vec y}_N) \,
\vert N \rangle_{MF} \nonumber\\
&=& \ {1 \over N!} \ \langle  O \vert e^{J({\vec y}_1)} \ \Phi({\vec y}_1)
\ e^{J({\vec y}_2)} \ \Phi({\vec y}_2) \ ..... \ e^{J({\vec y}_N)}
\ \Phi({\vec y}_N) \nonumber \\
& & \times \int \ d^2 x_1 \, \Pi({\vec x}_1) \ \int \ d^2 x_2 \,
\Pi({\vec x}_2) \ ...... \ \int \ d^2 x_N \, \Pi({\vec x}_N)
\vert O \rangle \ .
\eeqarr
Now we can use the identity,
\beq
\Phi({\vec y}_1) \,
e^{J({\vec y}_2)}   =   e^{J({\vec y}_2)} \, \Phi({\vec y}_1) \, (z_1-z_2)^m,
\eeq
to move all the factors of $e^J$ to the left, which yields
\beqarr
\psi_{MF} \ ( {\vec y}_1,{\vec y}_2, \ ..... \ {\vec y}_N) \nonumber \\
  \ &=& \ {1 \over N!} \prod_{i<j} (z_i - z_j)^m \  \ \langle O \vert
e^{ \sum_{i}  J({\vec y}_i)}
\,  \Phi({\vec y}_1) \ ....... \ \Phi({\vec y}_N)\nonumber\\
& & \times \int \ d^2 x_1 \, \Pi({\vec x}_1) \ \int \ d^2 x_2 \,
\Pi({\vec x}_2) \ ...... \ \int \ d^2 x_N \, \Pi({\vec x}_N)
\vert O \rangle \ .
\eeqarr
Next we use Wick's theorem for the product of the $\Phi$'s and $\Pi$'s and
note that the former annihilate the vacuum to obtain
\beqarr
\psi_{MF} \ ( {\vec y}_1,{\vec y}_2, \ ..... \ {\vec y}_N)
= \ \prod_{i<j} (z_i - z_j)^m \,
\langle O \vert e^{ \sum_{i}  J({\vec y}_i)}\,\vert O \rangle \ .
\eeqarr
Finally, as the vacuum has no particles, only the gaussian factor in $J$
contributes and we have the result,
\beq  \psi_{MF} \ ( {\vec y}_1,{\vec y}_2, \ ..... \ {\vec y}_N) \ = \
\prod_{i<j} (z_i - z_j)^m \ e^{ \sum_{i} {\mid z_i \mid ^2 \over 4l^2}}.
\eeq
Thus the mean field state directly yields the complete Laughlin wavefunction.
This is in contrast to the bosonized theory of ZHK, where the
mean field state contains only the correct phase of the Laughlin
wavefunction and not its zeroes or the gaussian factor; as already mentioned
those can be obtained correctly only upon including fluctuations.

\section{Summary and Prospects}
\label{summary}

In this paper we have set up a bosonic field theory for the
quantum Hall effect using an operator algebra based on Read's operator.
The field theory admits mean-field states at the fractions $\nu = 1/m$
that are ideal condensates in the Bose language and correspond exactly
to the Laughlin states in terms of the electrons. In order to treat
fluctuations about these states and to calculate the spectra of the
various collective modes it is necessary to perturb about the mean
field Hamiltonian. In our formulation, the mean-field Hamiltonian
has the simple form,
\beq
H_{MF} = - {\hbar^2 \over 2 \mu} \int d^2x \, \Pi({\vec x})\, \nabla^2
\, \Phi({\vec x})
\eeq
and hence its eigenstates are all known exactly. Nevertheless,
$H_{MF}$ is non-hermitian and hence states with different energies
are not necessarily orthogonal. (The full Hamiltonian is perfectly
hermitian; however the mean-field theory dictates that we decompose it
as the sum of two non-hermitian pieces.) This requires then, that the
perturbation theory explicitly take account of the non-orthogonality
and that we possess tractable expressions for the overlaps between different
states \cite{fn4}.
We expect to discuss progress on this problem elsewhere \cite{rs}.

\acknowledgements
We are grateful to M. P. Gelfand and F. D. M. Haldane for useful
discussions. This work was supported in part by NSF grants
Nos.\ DMR--9122385 and DMR--9157018 (SLS).


\begin{references}
\bibitem{fn5} In fact, there is a set of operators, each element of which
is long ranged in a different state. For linguistic simplicity we generally
refer to all the elements of the set in the singular, e.g. ``the bosonic
operator''.
\bibitem{fn1} The constraint Eq. (\ref{piphirelation}) relating $\Pi$ and
$\Phi$ does not need to be enforced separately; it is implicit in the equations
of motion for the two fields.
\bibitem{fn7} This conclusion follows as $w({\vec x},t)$ is harmonic and
hence does not allow transformations between field configurations with
different divergences.
\bibitem{fn2} There is also a related equation for $\Pi$.
\bibitem{fn6} We use the term ``mean-field solution'' loosely. Evidently,
what we present here are solutions to the classical equations of motion.
\bibitem{fn3} The off-diagonal correlation function is $ \Pi({\vec x})\,
\Phi({\vec x'})$ and it is long ranged in the mean field state.
\end{references}
\end{document}